\documentclass[aps,prl,twocolumn,preprintnumbers,superscriptaddress]{revtex4}

\usepackage{color}
\usepackage{multirow}
\usepackage[normalem]{ulem}

\usepackage{amsmath}
\usepackage{enumerate}
\usepackage{amsfonts}
\usepackage{epsfig}

\newcommand{\be}{\begin{equation}}
\newcommand{\ee}{\end{equation}}
\newcommand{\bea}{\begin{eqnarray}}
\newcommand{\eea}{\end{eqnarray}}

\def\le{\left}
\def\ri{\right}
\def\[{\left [}
\def\]{\right ]}
\def\nein{N_{\rm c}}
\def\nmax{N_{\rm f}}
\def\D{B}
\def\DD{C}
\def\DDD{D}
\def\DDDD{E}
\def\ka{k_{a}}
\def\fb{s_{\D}}
\def\GG{G_2}
\def\GN{G_{\rm N}}

\begin{document}

\title{Disordered Holographic Systems II:\\Marginal Relevance of Imperfection}

\preprint{MIT-CTP-4344}

\author{Allan Adams}
\affiliation{Center for Theoretical Physics, Massachusetts Institute of Technology,
Cambridge, MA 02139, USA
}
\author{Sho Yaida}
\affiliation{Center for Theoretical Physics, Massachusetts Institute of Technology,
Cambridge, MA 02139, USA
}

\date{January, 2012}

\begin{abstract}
We continue our study of quenched disorder in holographic systems, focusing on the effects of mild electric disorder.
By studying the renormalization group evolution of the disorder distribution at subleading order in perturbations away from the clean fixed point, we show that electric disorder is marginally relevant in $(2+1)$-dimensional holographic conformal field theories.
\end{abstract}
\maketitle

\newpage
The standard lore on disorder holds that, in the absence of an external magnetic field, all states are localized in $d\leq2+1$ spacetime dimensions, regardless of the strength of the disorder.
This was codified in a beautiful scaling argument by the gang of four~\cite{4gangs}.
For Fermi liquids, the scaling argument can be substantiated by explicit calculations of the beta-function for the conductance~\cite{electronic}.
However, when confronted with systems which cannot be described in terms of weakly interacting quasiparticles, such calculations are generally lacking.

An interesting test case for the standard lore is disorder in holographic quantum field theories~\cite{HolographyReviews}.
Despite having no quasiparticle descriptions, these models are computationally tractable thanks to a dual description via classical gravity in one higher dimension.
The extra dimension encodes the renormalization group scale of the field theory.
Quenched disorder can be implemented by imposing disordered boundary conditions for the classical fields in the emergent spacetime.
Tracing the renormalization group flow of the disorder distribution then reduces to solving the classical equations of motion of the dual gravitational system subject to disordered boundary conditions and studying how the effective distribution varies along the extra dimension.
This strategy was used in~\cite{1stAY} to show that the classical Harris criterion holds. In particular, Gaussian quenched electric disorder is marginal in two spatial dimensions at leading order in perturbations around the clean fixed point.

In this paper we show that quenched electric disorder in our holographic conformal field theory (CFT) is in fact marginally relevant in two spatial dimensions.
This is demonstrated by computing subleading corrections to the bulk energy-momentum tensor and the moments of the disorder distribution; resumming the resulting logarithms yields marginal relevance.
Our discussion will rely heavily on the formalism developed in~\cite{1stAY}, to which we refer the reader for details and further references.

As in~\cite{1stAY}, we focus on a $d$-dimensional strongly correlated CFT which is holographically dual to $(d+1)$-dimensional classical Einstein-Maxwell theory, whose equations of motion are,
\be
\frac{1}{\sqrt{|g|}}\partial_Q\le[\sqrt{|g|}g^{QP}g^{MN} F_{PN}\ri]=0\,,
\ee
\bea
R_{MN}-\frac{1}{2}R g_{MN}-\frac{d(d-1)}{2L^2}g_{MN}&&\\
=\frac{8\pi \GN}{g_{d+1}^2}\le[F_{MP}F_{N}^{\ P}-\frac{1}{4}F_{PQ}F^{PQ}g_{MN}\ri]&\equiv&{8\pi \GN}T_{MN}\,.\nonumber
\eea
The dimensionless constants $\frac{L^{d-1}}{\GN}\equiv \nein^2$ and $\frac{L^{d-3}}{g_{d+1}^2}\equiv \nmax^2$ are determined by the parameters of the boundary CFT.
We take a large-$\nein$ limit to ensure classicality of the bulk theory, but keep $\nmax\sim\nein$ to bring the role of gravitational backreaction to the fore.
Recalling that the chemical potential of the CFT is encoded by the boundary value of the electric potential in the bulk, we see that quenched electric disorder can be implemented by fixing the boundary value of $A_{0}(z; x^{0}, {\bf x})$ to take a time-independent random value, $V({\bf x})$, dictated by a suitable distribution $P_{V}\le[W({\bf x})\ri]$ over functions $W({\bf x})$~\cite{1stAY}.

In the limit of mild disorder, we can expand our bulk fields in perturbations around the clean background,
\bea
g_{MN}&=&g^{(0)}_{MN}+g^{(2)}_{MN}+g^{(4)}_{MN}+...\ \ \,, \\
A_{M}&=&A^{(1)}_{M}+A^{(3)}_{M}+....
\eea
Here, $g^{(0)}_{MN}$ is the unperturbed background metric, $A^{(1)}_{M}$ is the gauge field sourced by electric disorder potential, $V({\bf x})$, in the
ultraviolet, and $g^{(2n)}_{MN}$ and $A^{(2n+1)}_{M}$ for $n\geq1$ are generated by higher-order backreaction.
We fix gauge by putting the metric in the form,
\be
g_{MN}dx^Mdx^N=\frac{L^2}{z^2}dz^2+\sum_{\mu, \nu=0}^{d-1}g_{\mu\nu}dx^{\mu}dx^{\nu}.
\ee
and setting $A_{z}$=0.
In particular, $g^{(2n)}_{zM}=0$ for $n\geq1$.

We must also specify boundary conditions in the infrared and ultraviolet.
We require all field to be regular at the infrared horizon. In the ultraviolet, we may consider a simple model of quenched electric disorder~\cite{1stAY},
\be
\label{UVBC}
\lim_{z\rightarrow0}A_{\mu}=\delta_{0\mu}V_{\infty}({\bf x})
\ee
where $V_{\infty}({\bf x})$ is governed by the Gaussian disorder distribution functional
\bea
\label{QED}
P_{V_{\infty}}[W({\bf x})]
&=&N_{\sharp}e^{-\frac{1}{2f_{\rm dis}}\int d^{d-1}{\bf x}W({\bf x})^2}
\nonumber\\
&=&N_{\sharp}e^{-\frac{1}{2f_{\rm dis}}\int \frac{d^{d-1}{\bf k}}{(2\pi)^{d-1}}W({\bf k})W(-{\bf k})}.
\eea
However, this simple choice leads to a divergence in the disorder-averaged metric for $d\geq 2+1$~\cite{1stAY}.
To regulate this divergence we instead impose the following Dirichlet boundary condition on the hypersurface at $z=\frac{1}{\Lambda}$:
\bea
g_{\mu\nu}\Big|_{z=\frac{1}{\Lambda}}&=&{L^2\Lambda^2}\delta_{\mu\nu}\ \ \ {\rm and}\\
A_{\mu}\Big|_{z=\frac{1}{\Lambda}}&=&\delta_{0\mu}V^{\Lambda}({\bf x})
\eea
with the disorder distribution functional
\be
\label{cutQED}
P_{V_{\Lambda}}[W({\bf x})]=N'_{\sharp}e^{-\frac{1}{2f^{\Lambda}_{{\rm dis}}}\int \frac{d^{d-1}{\bf k}}{(2\pi)^{d-1}}W({\bf k})W(-{\bf k})e^{\frac{2|{\bf k}|}{\Lambda}}}.
\ee
Note that the $e^{\frac{2|{\bf k}|}{\Lambda}}$ softly cuts off high momentum modes.
As will be clear shortly [cf. Eqs.(\ref{background}) and (\ref{probe})], for $d=2+1$ at zero temperature, this is tantamount to choosing the condition (\ref{UVBC}) with $f_{{\rm dis}}=f^{\Lambda}_{\rm dis}$ for $A^{(1)}_{\mu}$ and setting $g^{(2n)}_{\mu\nu}\Big|_{z=\frac{1}{\Lambda}}=0$ and $A^{(2n+1)}_{\mu}\Big|_{z=\frac{1}{\Lambda}}=0$ for $n\geq1$.

We now solve the equations of motion order by order, focusing on the marginal case, $d=2+1$.
Working at zero temperature and in Euclidean time, $\tau=+ix^0$, the leading-order background is pure anti-de Sitter space,
\be
\label{background}
g^{(0)}_{MN} dx^M dx^N=\frac{L^2}{z^2}\le(dz^2+d\tau^2+\sum_{i=1}^{2}(dx^i)^2\ri).
\ee

At first order, the Maxwell equations give $A^{(1)}_i$=0 and
\be
\label{probe}
A^{(1)}_{\tau}({\bf k}; z)=(-i)V^{\Lambda}_{\infty}({\bf k})e^{-|{\bf k}|z}
\ee
where $V^{\Lambda}_{\infty}$ is governed by the functional (\ref{QED}) with $f_{\rm dis}=f^{\Lambda}_{\rm dis}$.
We see that it is regular at the Poincar\'{e} horizon ($z\rightarrow\infty$) and is governed by the functional (\ref{cutQED}) at $z=\frac{1}{\Lambda}$, with $\le(e^{-|{\bf k}|z}\ri)^2\Big|_{z=\frac{1}{\Lambda}}$ correctly reproducing the soft cutoff stipulated above.

At second order, $A^{(1)}$ generates an inhomogeneous energy-momentum tensor
\bea
8\pi\GN T^{(2)}_{MN}\le({\bf k}; z\ri)&=&-\frac{4\pi\GN z^2}{g_4^2L^2}\int\frac{d^2{\bf k'}}{4\pi^2}V_{\infty}^{\Lambda}\le({\bf k}-{\bf k'}\ri)V_{\infty}^{\Lambda}\le({\bf k'}\ri)\nonumber\\
&&\times t_{MN}({\bf k}-{\bf k'}, {\bf k'})e^{-\le(|{\bf k}-{\bf k'}|+|{\bf k'}|\ri)z}
\eea
with
\bea
t_{zz}\le({\bf k}_{1}, {\bf k}_{2}\ri)&=&|{\bf k}_{1}||{\bf k}_{2}|+{\bf k}_{1}\cdot{\bf k}_{2},\\
t_{zi}\le({\bf k}_{1}, {\bf k}_{2}\ri)&=&-i|{\bf k}_{1}|\le({\bf k}_{2}\ri)_{i}-i|{\bf k}_{2}|\le({\bf k}_{1}\ri)_{i},\\
t_{\tau\tau}\le({\bf k}_{1}, {\bf k}_{2}\ri)&=&|{\bf k}_{1}||{\bf k}_{2}|-{\bf k}_{1}\cdot{\bf k}_{2},,\ \ \ {\rm and}\\
t_{ij}\le({\bf k}_{1}, {\bf k}_{2}\ri)&=&\delta_{ij}\le(-|{\bf k}_{1}||{\bf k}_{2}|+{\bf k}_{1}\cdot{\bf k}_{2}\ri)\nonumber\\
&&-\le({\bf k}_{1}\ri)_{i}\le({\bf k}_{2}\ri)_{j}-\le({\bf k}_{2}\ri)_{i}\le({\bf k}_{1}\ri)_{j}\,.
\eea
This in turn activates $g^{(2)}_{\mu\nu}\le({\bf k}; z\ri)\equiv\frac{L^2}{z^2}\phi^{(2)}_{\mu\nu}\le({\bf k}; z\ri)$ through the Einstein equations.
There are two cases to be dealt with separately: ${\bf k}=0$ and ${\bf k}\ne0$.

For ${\bf k}=0$, the only relevant ingredient for our later calculations turns out to be the disorder-averaged metric. Solving the homogeneous Einstein equation with $[8\pi\GN T^{(2)}_{zz}]_{\rm d.a.}=0$, $[8\pi\GN T^{(2)}_{\tau\tau}]_{\rm d.a.}=-\frac{3f^{\Lambda}_{\rm dis}\GN}{2g_4^2L^2z^2}$, and $[8\pi\GN T^{(2)}_{ij}]_{\rm d.a.}=\frac{3f^{\Lambda}_{\rm dis}\GN}{4g_4^2L^2z^2}\delta_{ij}$ yields
\bea
\le[\phi^{(2)}_{\tau\tau}\ri]_{\rm d.a.}&=&-\le(\frac{f^{\Lambda}_{\rm dis}\GN}{g_4^2L^2}\ri){\rm log}\le(\Lambda z\ri)\ \ \ {\rm and}\\
\le[\phi^{(2)}_{ij}\ri]_{\rm d.a.}&=&+\frac{\delta_{ij}}{2}\le(\frac{f^{\Lambda}_{\rm dis}\GN}{g_4^2L^2}\ri){\rm log}\le(\Lambda z\ri).
\eea
The overall sign in front is crucial for marginal relevance of the quenched electric disorder.

For ${\bf k}\ne0$, solving the inhomogeneous linearized Einstein equation~\cite{trickA}, we obtain $\phi_{\tau i}=0$,
\bea
\phi^{(2)}_{\tau\tau}({\bf k}; z)&=&-\D({\bf k}; z)-\DD({\bf k}; z),\ \ \ {\rm and}\\
\phi^{(2)}_{ij}({\bf k}; z)&=&\le(-\delta_{ij}+3\frac{\le({\bf k}\ri)_{i}\le({\bf k}\ri)_{j}}{{\bf k}^2}\ri)\D({\bf k}; z)\\
&+&\le(\delta_{ij}-\frac{\le({\bf k}\ri)_{i}\le({\bf k}\ri)_{j}}{{\bf k}^2}\ri)\DD({\bf k}; z)\nonumber\\
&+&\le(\frac{\le({\bf k}\ri)_{i}\le({\bf k}\ri)_{j}}{{\bf k}^2}\ri)\DDD({\bf k}; z)\nonumber\\
&-&i\le({\bf k}\ri)_{i}\DDDD_{j}({\bf k}; z)-i\le({\bf k}\ri)_{j}\DDDD_{i}({\bf k}; z)\nonumber
\eea
with
\bea
\D({\bf k}; z)&=&-\int_{\frac{1}{\Lambda}}^{z}dz'\le[\frac{i\le({\bf k}\ri)_{i}}{{\bf k}^2}8\pi\GN T^{(2)}_{zi}({\bf k}; z')\ri],\nonumber\\
\DD({\bf k}; z)&=&-\GG\le(|{\bf k}|z\ri)\int_{\frac{1}{\Lambda}}^{z}dz'\frac{z'^2}{\GG^2\le(|{\bf k}|z'\ri)}\int_{z'}^{\infty}dz''\frac{\GG\le(|{\bf k}|z''\ri)}{z''^2}\nonumber\\
&&\times\le[8\pi\GN\le\{T^{(2)}_{\tau\tau}-T^{(2)}_{ii}+\frac{\le({\bf k}\ri)_{i}\le({\bf k}\ri)_{j}}{{\bf k}^2}T^{(2)}_{ij}\ri\}\ri]({\bf k}; z''),\nonumber\\
\DDD({\bf k}; z)&=&\le(-\frac{z^2}{2}+\frac{1}{2\Lambda^2}\ri)8\pi\GN T^{(2)}_{zz}\le({\bf k}; \frac{1}{\Lambda}\ri)\nonumber\\
&&-2\int_{\frac{1}{\Lambda}}^{z} dz' z'\int_{\frac{1}{\Lambda}}^{z'}dz''\frac{8\pi\GN T^{(2)}_{zz}\le({\bf k}; z''\ri)}{z''},\ \ \ {\rm and}\nonumber\\
\DDDD_i({\bf k}; z)&=&\int_{\frac{1}{\Lambda}}^{z}dz'\le[\frac{2}{{\bf k}^2}\le(\delta_{ij}-\frac{\le({\bf k}\ri)_{i}\le({\bf k}\ri)_{j}}{{\bf k}^2}\ri)8\pi\GN T^{(2)}_{zj}({\bf k}; z')\ri]\nonumber
\eea
where we defined $\GG\le(y\ri)\equiv\le(1+y\ri)e^{-y}$.

Finally, at third order, $A^{(3)}_i=0$ and, solving
\be
\le[\partial_z^2-{\bf k}^2\ri]A^{(3)}_{\tau}({\bf k}; z)=S^{(3)}({\bf k}; z)
\ee
with
\bea
S^{(3)}&\equiv&\frac{1}{2}\le\{\partial_z\le(\phi^{(2)}_{\tau\tau}-\phi^{(2)}_{jj}\ri)\ri\}\le(\partial_zA^{(1)}_{\tau}\ri)\nonumber\\
&&+\frac{1}{2}\le\{\partial_i\le(\phi^{(2)}_{\tau\tau}-\phi^{(2)}_{jj}\ri)\ri\}\le(\partial_iA^{(1)}_{\tau}\ri)\nonumber\\
&&+\partial_i\le\{\phi^{(2)}_{ij}\le(\partial_jA^{(1)}_{\tau}\ri)\ri\},
\eea
we get
\be
A^{(3)}_{\tau}=-e^{-|{\bf k}|z}\int_{\frac{1}{\Lambda}}^{z}dz'e^{2|{\bf k}|z'}\int_{z'}^{\infty}dz'' e^{-|{\bf k}|z''}S^{(3)}({\bf k}; z'')
\ee
which is regular at $z=\infty$ and vanishes at $z=\frac{1}{\Lambda}$.

To diagnose the relevance of the quenched electric disorder, we define
\bea
I_{\rm dis}\le(z\ri)&\equiv&\le[A_{\mu}({\bf x}; z)A^{\mu}({\bf x}; z)\ri]_{\rm d.a.}\\
&=&\int\frac{d^2{\bf k}}{4\pi^2}\int\frac{d^2{\bf k'}}{4\pi^2}\le[A_{\mu}({\bf k}; z)A^{\mu}({\bf k'}; z)\ri]_{\rm d.a.}
\eea
where we used the fact that disorder-averaged quantities are translationally invariant.
This characterizes the intensity of the disorder as a function of energy scale $\frac{1}{z}$.

At the leading order
\be
I^{(2)}_{\rm dis}\le(z\ri)=\int\frac{d^2{\bf k}}{4\pi^2}f^{\Lambda}_{\rm dis}\frac{z^2}{L^2}(-i)^2e^{-2|{\bf k}|z}=-\frac{f^{\Lambda}_{\rm dis}}{8\pi L^2},
\ee
the constancy of which reflects the marginality of the disorder at this order.
The same calculation for $d\ne2+1$ reproduces the Harris criterion, which has also been confirmed for the holographic CFT by computing disorder corrections to thermodynamic quantities~\cite{1stAY}.

At the next order we have
\bea
I^{(4)}_{\rm dis}\le(z\ri)&=&2\frac{z^2}{L^2}\le[A^{(1)}_{\tau}({\bf x}, z)A^{(3)}_{\tau}({\bf x}, z)\ri]_{\rm d.a.}\nonumber\\
&&-\frac{z^2}{L^2}\le[A^{(1)}_{\tau}({\bf x}, z)\phi^{(2)}_{\tau\tau}({\bf x}, z)A^{(1)}_{\tau}({\bf x}, z)\ri]_{\rm d.a.}\nonumber\\
&=&-\frac{f^{\Lambda}_{\rm dis}}{8\pi L^2}\le(\frac{f^{\Lambda}_{\rm dis}\GN}{g_4^2L^2}\ri)\times {\tilde I}^{(4)}\le(\Lambda z\ri).
\eea
A mathematical trick to extract the logarithmic term, the sign of which tells relevancy apart from irrelevancy, is to take $\Lambda\frac{\partial}{\partial \Lambda}I^{(4)}_{\rm dis}$ and send $\Lambda z\rightarrow \infty$, while pretending that $V^{\Lambda}_{\infty}$ are $\Lambda$-independent.

In evaluating $\Lambda\frac{\partial}{\partial \Lambda}I^{(4)}_{\rm dis}$, there are three different ways of contracting four $V^{\Lambda}_{\infty}$'s involved at this order.
First, we may disorder-average $g^{(2)}_{\mu\nu}$ and then disorder-average the remaining two factors of $A^{(1)}_{\mu}$~\cite{footA}.
Both averages are straightforward, leading to a contribution of the form,
\be
\Lambda\frac{\partial}{\partial \Lambda}{\tilde I}^{(4)}\Big|_{[g][AA]}=+\frac{3}{2}+O\le(\frac{{\rm log}\le(\Lambda z\ri)}{\Lambda z}\ri).
\ee
The other two contractions involve one factor of $V^{\Lambda}_{\infty}$ in $g^{(2)}_{\mu\nu}$ contracting with one of the $A^{(1)}_{\mu}$ and the other $V^{\Lambda}_{\infty}$ in $g^{(2)}_{\mu\nu}$ contracting with the other $A^{(1)}_{\mu}$. It turns out that neither contributes a logarithmic term, as we show explicitly in the Appendix.
All in all we find,
\be
I^{(4)}_{\rm dis}\le(z\ri)=-\frac{f^{\Lambda}_{\rm dis}}{8\pi L^2}\le(\frac{f^{\Lambda}_{\rm dis}\GN}{g_4^2L^2}\ri)\times\le[+\frac{3}{2}{\rm log}\le(\Lambda z\ri)+c_0+...\ri]
\ee
where $c_0$ is a constant of order $1$ and dots indicate the terms which asymptote to zero for $z\gg\frac{1}{\Lambda}$. We see that $I^{(2)}_{\rm dis}\le(z\ri)+I^{(4)}_{\rm dis}\le(z\ri)$ grows toward infrared. In other words, the disorder is marginally relevant, with an effective disorder strength
\be
f^{{\rm eff}}_{\rm dis}(z)=f^{\Lambda}_{\rm dis}+\le(f^{\Lambda}_{\rm dis}\ri)^2\frac{3}{2}\frac{\nmax}{\nein}{\rm log}\le(\Lambda z\ri).
\ee

We can check this result by studying the disorder-averaged bulk energy-momentum tensor, whose growth would indicate large corrections to the geometry in the infrared.
A similar, if more involved, analysis shows that we again receive a logarithmic contribution only from contractions involving $\le[g^{(2)}_{\mu\nu}\ri]_{\rm d.a.}$.
The results are~\cite{trickB}:
\bea
&&[8\pi\GN T^{(4)z}_{\ \ \ \ z}]_{\rm d.a.}=\frac{-9}{4L^2}\le(\frac{f^{\Lambda}_{\rm dis}\GN}{g_4^2L^2}\ri)^2\le[c_1+...\ri];\nonumber\\
&&[8\pi\GN T^{(4)\tau}_{\ \ \ \ \tau}]_{\rm d.a.}=\frac{-9}{4L^2}\le(\frac{f^{\Lambda}_{\rm dis}\GN}{g_4^2L^2}\ri)^2\le[{\rm log}\le(\Lambda z\ri)+c_2+...\ri];\nonumber\\
&&[8\pi\GN T^{(4)i}_{\ \ \ \ j}]_{\rm d.a.}=\frac{+9\delta^{i}_{\ j}}{8L^2}\le(\frac{f^{\Lambda}_{\rm dis}\GN}{g_4^2L^2}\ri)^2\le[{\rm log}\le(\Lambda z\ri)+c_3+...\ri].\nonumber
\eea
The $c_{i}$'s are constants of order $1$ satisfying $c_1+c_2-c_3=0$. The absence of the logarithmic term in $[8\pi\GN T^{(4)z}_{\ \ \ \ z}]_{\rm d.a.}$ is consistent with $[8\pi\GN T^{(2)z}_{\ \ \ \ z}]_{\rm d.a.}=0$, while comparison with $[8\pi\GN T^{(2)\mu}_{\ \ \ \ \nu}]_{\rm d.a.}$ once again reveals the ``golden number," $+\frac{3}{2}\frac{\nmax}{\nein}$, which adds confidence to the claim that we are looking at right quantities to diagnose relevance of the disorder.


We conclude that quenched electric disorder is marginally relevant around the clean fixed point of our $(2+1)$-dimensional holographic CFT.
Together with the Harris criterion found at leading order in~\cite{1stAY}, we see that the standard lore is consistent with our results for mild electric disorder in the holographic setting: quenched electric disorder is irrelevant for $d>2+1$, relevant for $d<2+1$, and marginally relevant for $d=2+1$.

Given this close parallel with weakly-coupled disordered electronic systems around their clean fixed points~\cite{electronic}~\cite{caveat}, it is natural to wonder whether localization is universal in holographic CFTs with strong electric disorder in arbitrary spacetime dimension.
If so, there must be an unstable disordered critical point governing a metal-insulator transition in holographic CFTs in $d>2+1$ spacetime dimensions, with the insulating phase dual to a strongly inhomogeneous black hole horizon. We will return to this question in future work.

We thank Stephen Shenker for helpful discussions and Omid Saremi for collaboration on related questions.
The research of A.A. is supported by the DOE under contract \#DE-FC02-94ER40818.
S.Y. is supported by a JSPS Postdoctoral Fellowship for Research Abroad.  

\appendix

\section{APPENDIX}
\label{cross}
Let us explicitly work out the contribution to the cross-contracted piece $\Lambda\frac{\partial}{\partial \Lambda}{\tilde I}^{(4)}\Big|_{\rm cross}$ coming from $\D({\bf k}; z)$.
From $-\frac{z^2}{L^2}\le[A^{(1)}_{\tau}\le(\Lambda\frac{\partial}{\partial \Lambda}\phi^{(2)}_{\tau\tau}\ri)A^{(1)}_{\tau}\ri]_{\rm d.a.}$, we receive
\bea
&-&\frac{\le(f^{\Lambda}_{\rm dis}\ri)^2}{L^2}\le(\frac{8\pi\GN}{g_4^2L^2}\ri)\frac{z^2}{\Lambda^3}\int\frac{d^2{\bf k}_{1}}{4\pi^2}\int\frac{d^2{\bf k}_{2}}{4\pi^2}\nonumber\\
&&e^{-\le(|{\bf k}_{1}|+|{\bf k}_{2}|\ri)\le(z+\frac{1}{\Lambda}\ri)}\le[\frac{\le(|{\bf k}_{1}||{\bf k}_{2}|+{\bf k}_{1}\cdot{\bf k}_{2}\ri)}{\le({\bf k}_{1}+{\bf k}_{2}\ri)^2}\le(|{\bf k}_{1}|+|{\bf k}_{2}|\ri)\ri].\nonumber
\eea
After rescaling to ${\tilde{\bf k}}_{i}\equiv z{\bf k}_{i}$, we see that this term is of order $O(\frac{1}{\Lambda^3z^3})$. Note that there is no singularity from $\frac{1}{\le({\bf k}_{1}+{\bf k}_{2}\ri)^2}$, thanks to the numerator $\le(|{\bf k}_{1}||{\bf k}_{2}|+{\bf k}_{1}\cdot{\bf k}_{2}\ri)$.

As for terms involving $A^{(3)}_{\tau}$, note that
\bea
&&\Lambda\frac{\partial}{\partial \Lambda}A^{(3)}_{\tau}=-e^{-|{\bf k}|z+|{\bf k}|\frac{2}{\Lambda}}\frac{1}{\Lambda}\int_{\frac{1}{\Lambda}}^{\infty}dz'e^{-|{\bf k}|z'}S^{(3)}({\bf k}; z')\nonumber\\
&&\ \ \ \ \ -e^{-|{\bf k}|z}\int_{\frac{1}{\Lambda}}^{z}dz'e^{2|{\bf k}|z'}\int_{z'}^{\infty}dz'' e^{-|{\bf k}|z''}\Lambda\frac{\partial S^{(3)}}{\partial \Lambda}({\bf k}; z'')\nonumber
\eea
The first term makes contribution to $\Lambda\frac{\partial}{\partial \Lambda}{\tilde I}^{(4)}$ of the form
\bea
\frac{-8}{\pi^2}\int d^2{\bf k}_{1}d^2{\bf k}_{2}\le[\frac{|{\bf k}_{1}||{\bf k}_{2}|+{\bf k}_{1}\cdot{\bf k}_{2}}{\le({\bf k}_{1}+{\bf k}_{2}\ri)^2}\ri]\le(\frac{z^2}{\Lambda}\ri)e^{-2|{\bf k}_{1}|z-2|{\bf k}_{2}|\frac{1}{\Lambda}}\nonumber\\
\le[\le(\frac{|{\bf k}_{2}|}{4}+\frac{\fb\le({\bf k}_{1}, {\bf k}_{2}\ri)}{4\ka}\ri)\le(\frac{2}{\Lambda^2}+\frac{2}{\ka \Lambda}+\frac{1}{\ka^2}\ri)\ri]\nonumber
\eea
where we defined $\ka\equiv|{\bf k}_{1}|+|{\bf k}_{2}|$ and $\fb\le({\bf k}_{1}, {\bf k}_{2}\ri)\equiv-|{\bf k}_{2}|^2-2{\bf k}_{1}\cdot{\bf k}_{2}+\frac{3\le\{\le({\bf k}_{1}+{\bf k}_{2}\ri)\cdot{\bf k}_{1}\ri\}\le\{\le({\bf k}_{1}+{\bf k}_{2}\ri)\cdot{\bf k}_{2}\ri\}}{\le({\bf k}_{1}+{\bf k}_{2}\ri)^2}$. This time, rescaling to ${\check{\bf k}}_{1}\equiv z{\bf k}_{1}$ and ${\check{\bf k}}_{2}\equiv \frac{{\bf k}_{2}}{\Lambda}$, one can argue that it is of order $O\le(\frac{1}{\Lambda z}\ri)$.
The other term coming from $A^{(3)}_{\tau}$ contributes
\bea
\frac{8}{\pi^2}\int d^2{\bf k}_{1}d^2{\bf k}_{2}\le[\frac{|{\bf k}_{1}||{\bf k}_{2}|+{\bf k}_{1}\cdot{\bf k}_{2}}{\le({\bf k}_{1}+{\bf k}_{2}\ri)^2}\fb\le({\bf k}_{1}, {\bf k}_{2}\ri)\ri]\le(\frac{z^2}{\Lambda^3}\ri)\nonumber\\
\le[\frac{\le(e^{-\le(|{\bf k}_{1}|+|{\bf k}_{2}|\ri)\le(z+\frac{1}{\Lambda}\ri)}-e^{-2|{\bf k}_{1}|z-2|{\bf k}_{2}|\frac{1}{\Lambda}}\ri)}{|{\bf k}_{2}|-|{\bf k}_{1}|}\ri]\nonumber.
\eea
Note that there is no singularity from $|{\bf k}_{1}|\rightarrow|{\bf k}_{2}|$ as $e^{-\le(|{\bf k}_{1}|+|{\bf k}_{2}|\ri)\le(z+\frac{1}{\Lambda}\ri)}$ and $e^{-2|{\bf k}_{1}|z-2|{\bf k}_{2}|\frac{1}{\Lambda}}$ approach each other. Then, the term involving $e^{-\le(|{\bf k}_{1}|+|{\bf k}_{2}|\ri)\le(z+\frac{1}{\Lambda}\ri)}$, upon rescaling to ${\tilde{\bf k}}_{i}$, gives the contribution of order $O(\frac{1}{\Lambda^3z^3})$ while the term involving $e^{-2|{\bf k}_{1}|z-2|{\bf k}_{2}|\frac{1}{\Lambda}}$, upon rescaling to ${\check{\bf k}}_{i}$ gives the contribution of order $O\le(\frac{1}{\Lambda z}\ri)$.

The cross-contracted terms involving $\D({\bf k}; z)$ thus do not contribute any logarithms.
Similar logic applies to the $\DDD({\bf k}; z)$ and $\DDDD_{i}({\bf k}; z)$ contributions.  For $\DD({\bf k}; z)$ it turns out to be easier to simply show that setting $\frac{1}{\Lambda}=0$ yields a finite contribution to ${\tilde I}^{(4)}$, which in particular implies the absence of any divergent logarithms.


\begin{thebibliography}{11}
\bibitem{footA}
  Note that $A^{(3)}$ is essentially a product of $g^{(2)}_{\mu\nu}$ and $A^{(1)}_{\mu}$.
\bibitem{trickB}
  A quick path to get $\le[T^{(4)z}_{\ \ \ \ z}\ri]_{\rm d.a.}$ is to use $T^M_{\ \ \ M}=0$ for $d=2+1$ and the energy-momentum conservation $\nabla_MT^M_{\ \ \ z}=0$, together with the regularity at the horizon. Incidentally, the same logic implies that $\le[T^{(2)z}_{\ \ \ \ z}\ri]_{\rm d.a.}=0$.
\bibitem{caveat}
  One may be concerned with the fact that the $U(1)$ symmetry of the holographic CFT, whose current is coupled to the random potential, is global on the field theory side. Although we do not fully justify it, this may be a red herring: at least, for electronic systems, one can often proceed without explicitly taking $U(1)$ gauge field into account. For holographic systems, the literature on the issue of gauging $U(1)$ symmetry is surprisingly scarce and it calls for careful development.
\end{thebibliography}
\end{document}